\documentstyle[epsfig]{mn}
\title[Peculiar Velocities of Galaxy Clusters]
      {Peculiar Velocities of Galaxy Clusters}
\author[J.M.\ Colberg et al.]
  {J.M.\ Colberg,$^1$
   S.D.M.\ White,$^1$ T.J.\ MacFarland,$^{2,3}$ A.\ Jenkins,$^4$ \newauthor 
   F.R.\ Pearce,$^4$ C.S.\ Frenk,$^4$ P.A.\ Thomas,$^5$ 
   H.M.P.\ Couchman$^6$\\
   $^1$Max--Planck--Institut f\"ur Astrophysik,
       Karl--Schwarzschild--Str.\ 1, D--85740 Garching, Germany\\
   $^2$Rechenzentrum Garching, Boltzmannstr.\ 2, D--85740 Garching,
      Germany\\
   $^3$now at Enterprise Architecture Group, Global Securities
      Industry Group, 1185 Avenue of the Americas, New York 10036, USA\\
   $^4$Physics Dept., University of Durham, Durham DH1 3LE, UK\\
   $^5$CEPS, University of Sussex, Brighton BN1 9QH, UK\\
   $^6$Dept.\ of Astronomy, University of Western Ontario, London,
       Ontario N6A 3K7, Canada}
\begin{document}
\newcommand{\etal}{et al.}
\maketitle
\begin{abstract}
We investigate the peculiar velocities predicted for galaxy clusters
by theories in the cold dark matter family. A widely used hypothesis
identifies rich clusters with high peaks of a suitably smoothed
version of the linear density fluctuation field. Their peculiar 
velocities are then obtained by extrapolating the
similarly smoothed linear peculiar velocities at the positions 
of these peaks. We test these ideas using large high resolution
N--body simulations carried out within the Virgo supercomputing 
consortium. We find that at early times the barycentre of the 
material which ends up in a rich cluster is generally very close to
a high peak of the initial density field. Furthermore
the mean peculiar velocity of this material agrees well
with the linear value at the peak. The late-time growth of
peculiar velocities is, however, systematically underestimated
by linear theory. At the time clusters are identified we find
their {\it rms} peculiar velocity to be about 40\% larger than
predicted. Nonlinear effects are particularly important 
in superclusters. These systematics must be borne in mind when 
using cluster peculiar velocities to estimate the parameter combination
$\sigma_8\Omega^{0.6}$. 
\end{abstract}
\section{Introduction} \label{intro}
The motions of galaxy clusters are thought to result from
gravitational forces acting over the very large scales on which
superclusters are assembled. The {\it rms} deviations from uniformity
on such scales appear to be small, and so may be adequately described
by the linear theory of fluctuation growth. For a linear density field
of given power spectrum the {\it rms} peculiar velocity
is proportional to $\sigma_8\Omega^{0.6}$ where $\Omega$ is the cosmic
density parameter and $\sigma_8$, the {\it rms} mass fluctuation in a
sphere of radius $8~h^{-1}$Mpc, is a conventional measure of the
amplitude of fluctuations (e.g. Peebles 1993). (As usual the Hubble 
constant is
expressed as $H_0 = 100\,h\,$km/sec/Mpc.)   Distance indicators
such as the Tully-Fisher or $D_n$--$\sigma$ relations allow the
peculiar velocities of clusters to be measured, thus providing
a direct estimate of this parameter combination (see, for example, 
Strauss \& Willick 1996). 

Essentially the same parameter combination can also be estimated from the 
{\it abundance} of galaxy clusters (e.g. White et al.\
1993) and a comparison of the two estimates could in principle provide
a check on the shape of the assumed power spectrum and on the assumption
that the initial density field had gaussian statistics. In practice
this is difficult because of the uncertainties in relating observed 
cluster samples to the objects for which quantities are calculated 
in linear theory or measured from N-body simulations. The standard
linear model was introduced by Bardeen et al.\ (1986; hereafter BBKS). 
It assumes that clusters can be identified with ``sufficiently'' high 
peaks of the linear density field after convolution with a
``suitable'' smoothing kernel. The peculiar velocity of a cluster 
is identified with the linear peculiar velocity of the corresponding 
peak extrapolated to the present day. In the present paper we study
the limitations both of this model and of direct N-body simulations 
by comparing their predictions for clusters on a case by case basis.

In the next section we summarize both the linear predictions for the growth of
peculiar velocities and the BBKS formulae for the values expected
at peaks of the smoothed density field. Section 3 then presents our
set of N-body simulations and outlines our procedures for identifying
peaks in the initial conditions and clusters at $z=0$.
Section 4 begins by studying how well peaks correspond to
the initial barycentres of clusters; we then show that
the smoothed linear velocity at a peak agrees well with the mean
linear velocity of its cluster; finally we show that the
growth of cluster peculiar velocities is systematically stronger at
late times than linear theory predicts. A final section presents a brief 
discussion of these results.

%%%%%%%%%%%%%%%%%%%%%%%%%%%%%%%%%%%%%%%%%%%%%%%%%%%%%%%%%%%%%%%%%%%%%

\section{Linear Predictions for the Peculiar Velocities of Peaks}

\subsection{The Growth of Peculiar Velocities}
\label{pvlt}

According to the linear theory of gravitational instability in a dust
universe \cite{Peebles93}, the peculiar velocity of every mass element
grows with cosmic scale factor $a$  as\begin{equation}
v\propto a \dot{D}\,,
\label{scale}
\end{equation}
where $a(t)$ is obtained from the
Friedman equation
\begin{equation}
\left(\frac{\dot{a}}{a}\right)^2 = \Omega_0\,a^{-3} + 
          (1-\Omega_0-\Lambda_0)\,a^{-2} + \Lambda_0\,,
\label{friedman}
\end{equation}
and $D(t)$ is the growth factor for linear density perturbations, 
$\delta({\bf x},t) = D(t) \delta_0({\bf x})$. 
$\Omega_0$ and $\Lambda_0$ are the density parameter and the
cosmological constant at $z=0$, respectively, and we define
$a=1$ at this time. A number of accurate approximate forms are known for 
the relations between $D$ and $a$ and can be used to cast the scaling
of eq.\ (\ref{scale}) into a more convenient form.  We write
\begin{equation}
\dot{D} \equiv \frac{{\rm d}D}{{\rm d}t} = \frac{{\rm d}D}{{\rm d}a}
\frac{{\rm d}a}{{\rm d}t}\,,
\label{dDdt}
\end{equation}
and substitute for ${\rm d}a/{\rm d}t$ from the Friedman equation
(\ref{friedman}). Lahav \etal{} (1991) give an approximation for
${\rm d}D/{\rm d}a$ in the combination
\begin{equation}
f(a) \equiv \frac{{\rm d}D}{{\rm d}a}\frac{a}{D} \approx \left( 
   \frac{\Omega_0 a^{-3}}{\Omega_0\,a^{-3}+(1-\Omega_0-\Lambda_0)\,a^{-2}+\Lambda_0} 
   \right)^{0.6}\,.
\label{lahav}
\end{equation}
For $a=1$ this gives the standard factor $f\approx \Omega_0^{0.6}$ 
which appears when predicting the peculiar velocities produced by a
given overdensity field. Carroll et al. (1992) used this result to
derive an approximation for $D(a)$ itself,
\begin{equation}
D \approx a\,g(a)\,,
\label{carroll}
\end{equation}
where
\begin{equation}
g(a) = \frac{5}{2}\frac{\Omega(a)}
   {\Omega^{4/7}(a) -
   \Lambda(a) + \left( 1 + \frac{\Omega(a)}{2}\right)
   \left(1+\frac{\Lambda(a)}{70}\right)}
\end{equation}
with 
\begin{eqnarray}
\Omega(a) & = & \frac{\Omega_0}
                     {a+\Omega_0(1-a)+\Lambda(a^3-a)}\,, \\
\Lambda(a) & = & \frac{\Lambda a^3}
                     {a+\Omega_0(1-a)+\Lambda(a^3-a)}\,. 
\end{eqnarray}
Combining these equations, we obtain an explicit approximation for 
the growth of peculiar velocities,
\begin{equation}
v\propto f(a)\,g(a)\,a^2
\sqrt{\Omega_0\,a^{-3}+(1-\Omega_0-\Lambda_0)\,a^{-2}+\Lambda_0}\,. 
\label{scaling}
\end{equation}
For the simple Einstein-de Sitter case where $\Omega_0=1$ and
$\Lambda=0$, these formulae reduce to the exact results
$D=a\propto t^{2/3}$ and $v\propto \sqrt{a}$.

Recently, Eisenstein (1997) has shown that the exact solutions for 
$D$ and $f(a)$ can be given explicitly in terms of elliptic
integrals. He also shows that the above approximations always have 
fractional errors 
better than 2\% if $\Omega>0.1$. We therefore work with the simpler
approximate formulae in the present paper.

%%%%%%%%%%%%%%%%%%%%%%%%%%%%%%%%%%%%%%%%%%%%%%%%%%%%%%

\subsection{The Velocities of Peaks}
\label{peakvel}

The idea that the statistical
properties of nonlinear objects like galaxy clusters can be inferred
from the initial linear density field was developed in considerable 
detail in the monumental paper of Bardeen et al. (1986; BBKS). If
the initial fluctuations are assumed to be a gaussian random field, 
they are specified completely by their power spectrum, $P(k)$.
Similarly, any smoothed version of this initial field is
specified completely by its own power spectrum, $P(k)W^2(kR)$, where
$W(kR)$ is the Fourier transform of the spherical smoothing kernel
and $R$ is a measure of its characteristic radius. In particular, 
BBKS showed how the abundance and {\it rms} peculiar velocity of peaks 
of given height can be expressed 
in terms of integrals over $P(k)W^2(kR)$. The difficulty in
connecting this model with real clusters lies in the ambiguity in
deciding what smoothing kernel, characteristic scale, and peak height 
are appropriate. Typically the smoothing kernel is taken to
be a gaussian or a top-hat, $R$ is chosen so that the kernel contains 
a mass similar to the minimum mass of the cluster sample, and the 
height is assumed sufficient for a spherical perturbation to collapse
by $z=0$.

The smoothed initial peculiar velocity field is isotropic and gaussian
with a three-dimensional dispersion given by
\begin{equation}
\sigma_v(R) \equiv H\,\Omega^{0.6}\,\sigma_{-1}(R)\,,
\end{equation}
where, in the notation of BBKS, $\sigma_{j}$ is defined for any
integer $j$ by
\begin{equation}
\sigma_j^2(R) = \frac{1}{2\pi^2} \int P(k)\,W^2(kR)\, k^{2j+2}\, {\rm d}k\,.
\label{sigmaj}
\end{equation}
The {\it rms} peculiar velocity at peaks of the smoothed density field
differs systematically from $\sigma_v$; BBKS show that it is given by
\begin{equation}
\sigma_p(R) = \sigma_v(R) \sqrt{1-\sigma_0^4/\sigma_1^2\sigma^2_{-1}}\,.
\end{equation}
Note that this expression does not depend on the height of the
peaks. As shown in BBKS, the velocities of peaks are {\em statistically\/}
independent of their height.

Throughout this paper we will approximate the power spectra of CDM models
by the parametric expression of Bond \& Efstathiou (1984),
\begin{equation}
P(k, \Gamma) = \frac{A k}
           {\{1+[ak/\Gamma+(bk/\Gamma)^{3/2}+(ck/\Gamma)^2]^{\nu}\}^{2/\nu}}\,,
\label{power}
\end{equation}
where $a=(6.4/\Gamma)\,h^{-1}\,$Mpc, $b=(3.0/\Gamma)\,h^{-1}\,$Mpc,
$c=(1.7/\Gamma)\,h^{-1}\,$Mpc, $\nu=1.13$, and the shape parameter
$\Gamma$ is given for the models discussed below by
\begin{equation}
\Gamma = \left\{ \begin{array}{ll}
                 \Omega_0 h/[0.861+3.8(m_{10}^2\tau_{\rm d})^{2/3}]^{1/2} & 
                        \mbox{for $\tau$CDM,} \\
                 \Omega_0 h & \mbox{otherwise.}  
                 \end{array}
         \right.
\label{gamma}
\end{equation}
In the $\tau$CDM case $m_{10}$ is the $\tau$-neutrino mass in units of
10\,keV and $\tau_{\rm d}$ is its lifetime in years (White et al.\
1995). A detailed investigation of this model can be found in
Bharadwaj \& Sethi (1998). For the cosmologies used here, we take
the values shown in Table 1. Detailed calculations of the
power spectrum are actually better fit by slightly smaller values of 
$\Gamma$ than we assume (Sugiyama 1995).

The normalisation constant in equation (\ref{power}) can be
related to the conventional normalisation $\sigma_8$ by 
noting that $\sigma_8\equiv \sigma_0(8h^{-1}{\rm Mpc})$ and using
equation (\ref{sigmaj}) with a top-hat window function, $W_{TH}(x) =
3(x\sin x - \cos x)/x^3$. This corresponds to the linear 
fluctuation amplitude extrapolated to $z=0$ and can be
matched to observation by fitting either to the cosmic microwave
background fluctuations measured by COBE or to the observed abundance
of rich galaxy clusters. The models of this paper are
normalised using the second method (c.f.\ Eke et al.\ 1996), 
as reflected by the $\sigma_8$
values given in Table 1 together with the other parameters defining
the models.

In the following linear density fields are smoothed either
with a top-hat or with a gaussian. In the latter case the window
function is $W_G(x)=\exp[-x^2/2]$. It is unclear for either filter how
$R$ should be chosen in order to optimize the
correspondance between peaks and clusters. We follow previous practice
in assuming that cluster samples contain all objects with mass
exceeding some threshold $M_{min}$, and then choosing $R$ so that the
filter contains $M_{min}$. Hence $M_{min}=4\pi \bar{\rho} R^3/3$ in
the top-hat case and $M_{min}=(2\pi)^{3/2} \bar{\rho} R^3$ in the
gaussian case. The simulations analysed here have $\Omega_0 = 0.3$ or
1.0, and we will isolate cluster samples limited at
$M_{min}=3.5\times 10^{14} h^{-1}\mbox{M}_{\odot}$, the value
appropriate for Abell clusters of richness one and greater (e.g. White
et al. 1993). A detailed discussion of filtering schemes can be found in
Monaco (1998) and references therein.

Table 2 gives characteristic filter radii $R$ and values of 
$\sigma_v$ and $\sigma_p$ from equations (11) and (13) for both 
smoothings and for all the cosmological models we consider in this 
paper; the velocity dispersions are extrapolated to the linear values 
predicted at  $z=0$. The difference between $\sigma_v$ and $\sigma_p$ 
has often been ignored in the literature when predicting the peculiar 
velocities of galaxy clusters (e.g. Croft \& Efstathiou 1994; Bahcall 
\& Oh 1996; Borgani et al.\ 1997); 
for our models the two differ by about 15\%. Notice 
also that with our choice of filter radii, gaussian smoothing predicts
{\it rms} peculiar velocities about 10\% smaller than top-hat smoothing.

%%%%%%%%%%%%%%%%%%%%%%%%%%%%%%%%%%%%%%%%%%%%%%%%%%%%%%%%%%%%%%%%%%%%%%%%%%%%%
%

\section{The Simulations}

\subsection{The Code} \label{code}

The Virgo Consortium was formed in order to study the evolution of
structure and the formation of galaxies using
the latest generation of parallel supercomputers 
\cite{Jenkins96}. The code used  for the simulations of this paper
is called {\sevensize HYDRA}. The original serial code was developed by 
Couchman \etal{} (1995) and was parallelized for CRAY T3D's by Pearce 
et al.\ (1995) (A detailed description can be found in Pearce
\& Couchman (1997). T3D--{\sevensize HYDRA} is a parallel adaptive
particle-particle/particle-mesh (AP$^3$M) code implemented in CRAFT, 
a directive-based parallel Fortran developed by CRAY. 
It supplements the standard P$^3$M algorithm (Efstathiou \etal{} 1985)
by recursively placing 
higher resolution meshes, {\em refinements\/}, over heavily clustered
regions.  Refinements containing more than $\sim 10^5$ particles are
executed in parallel by all processors; smaller refinements 
are completed using a task farm approach. This T3D version currently
includes an SPH treatment of gas dynamics, but this was not used for
the simulations of this paper.

A second version of {\sevensize HYDRA}, based on CRAY's shared memory
and message passing architecture, has been written by MacFarland 
et al.\ (1997). This can run on CRAY T3E's but does not
currently include refinement placing.

The simulations used here were run on the Cray T3D and T3E
supercomputers at the
computer center of the Max Planck Society in Garching and at
the Edinburgh Parallel Computing Centre.

\subsection{The Simulation Set} \label{simulations}

\begin{table}
\caption{The Virgo models}
\begin{center}
\begin{tabular}{lcccccc}
Model & $\Omega$ & $\Lambda$ & $h$ & $\sigma_8$ & $\Gamma$ & $z_{Start}$\\
OCDM & 0.3 & 0.0 & 0.7 & 0.85 & 0.21 & 119\\
$\Lambda$CDM & 0.3 & 0.7 & 0.7 & 0.90 & 0.21 & 30\\
SCDM & 1.0 & 0.0 & 0.5 & 0.51 & 0.50 & 35\\
$\tau$CDM & 1.0 & 0.0 & 0.5 & 0.51 & 0.21 & 35\\
\end{tabular}
\end{center}
\end{table}

A set of four matched N--body simulations of CDM universes was
completed in early 1997. Each follows
the evolution of structure within a cubic region $240h^{-1}$Mpc
on a side using $256^3$ equal mass particles and a gravitational
softening of $30h^{-1}$kpc. The choices of cosmological parameters
correspond to standard CDM (SCDM), to an Einstein-de Sitter model 
with an additional relativistic component ($\tau$CDM), to an open CDM model 
(OCDM), and to a flat low density model with a cosmological
constant ($\Lambda$CDM). A list of the parameters defining these
models is given in Table 1. 

In all models the initial fluctuation amplitude, and so the
value of $\sigma_8$, was set by
requiring that the models should reproduce the observed
abundance of rich clusters. Further details of this choice and of
other aspects of the simulations can be found in Jenkins et al.\
(1998). Note that each Fourier component of the initial fluctuation 
field had the same {\it phase} in each of these four simulations. As a
result there is an almost perfect correspondance between the clusters
in the four models. 

Because of their finite volume, these simulations contain no power
at wavelengths longer than $240h^{-1}$Mpc. Furthermore,
Fourier space is sampled quite coarsely on the largest scales for
which they do contain power, and so realisation to realisation
fluctuations on these scales can be significant. 
% In order to check for
% this we carried out a second simulation of the $\tau$CDM model with 
% different initial phases. 
The size of the effects can be judged from
Table 2 where we list the values of $\sigma_v$ and $\sigma_p$ obtained
for each model when the theoretical power spectrum is replaced in
equations (11) and (13) by the initial power spectrum of the model 
itself. These are systematically smaller than the values found before.
The difference is primarily a reflection of the loss of large-scale 
power.

%%%%%%%%%%%%%%%%%%%%%%%%%%%%%%%%%%%%%%%%%%%%%%%%%%%%%%%%%%%%%%%%%%%%%%%%%%%%%%%%

\begin{table*}
\caption{For each of the models, the following quantities are given:
the radius $R$ (second and fifth column) of the filter 
used in eq.\ (12); the three--dimensional velocity dispersions
$\sigma_v$ and $\sigma_p$ (third, fourth, sixth, and seventh column) 
obtained using eq.s (11) and (13) with the given filter
radii; the three--dimensional velocity dispersions
$\sigma_v$ and $\sigma_p$ (eighth, ninth, eleventh, and twelvth
column) obtained using eq.s (11) and (13) with the given filter
radii and the power spectra of the 
simulations themselves; the $rms$ linear overdensity $\Delta$ (tenth and
thirteenth column) smoothed with the given filter radii 
and extrapolated to $z=0$; the number of clusters $N_{\rm Cl}$
(fourteenth column) found in the simulations at $z=0$; the
three--dimensional velocity dispersions of peaks (fifteenth and
sixteenth column) in the initial conditions of the simulations
using the given filters; the three--dimensional linear velocity 
dispersions of clusters extrapolated to $z=0$;
and the three--dimensional measured velocity dispersion of 
clusters at $z=0$. The radii are given in
Mpc/$h$, the velocity dispersions in km/sec. Top Hat and Gaussian
filters are abbreviated as TH and G, respectively.}
\begin{center}
\begin{tabular}{lccccccccccccccccc}
 & \multicolumn{3}{c}{Top--Hat} & \multicolumn{3}{c}{Gaussian} &
\multicolumn{3}{c}{Sim TH} & \multicolumn{3}{c}{Sim Gauss} &
 & TH & G & Sim & Sim \\
 & \multicolumn{3}{c}{$\overbrace{\hspace{2.1cm}}$} & 
   \multicolumn{3}{c}{$\overbrace{\hspace{2.1cm}}$} &
 \multicolumn{3}{c}{$\overbrace{\hspace{2.3cm}}$} & 
 \multicolumn{3}{c}{$\overbrace{\hspace{2.3cm}}$} &
 & & & \\
\multicolumn{1}{c}{(1)} & (2) & (3) & (4) & (5) & (6) & (7) &
(8) & (9) & (10) & (11) & (12) & (13) &
(14) & (15) & (16)
& (17) & (18)\\
Model & $R$ & $\sigma_v$ & $\sigma_p$ & $R$ & $\sigma_v$ & $\sigma_p$ &
$\sigma_v$ & $\sigma_p$ & $\Delta$ & $\sigma_v$ & $\sigma_p$ & $\Delta$ &
$N_{\rm Cl}$ & $\sigma_{\rm Peak}$ & $\sigma_{\rm Peak}$
& $\sigma_{\rm lin}$ & $\sigma_{\rm z=0}$\\
OCDM         & 10.3 & 390 & 349 & 6.6 & 366 & 315 & 351 & 300 & 0.94
  & 321 & 258 & 0.96 & 62 & 253 & 266 & 280 & 407 \\
$\Lambda$CDM & 10.3 & 413 & 370 & 6.6 & 387 & 334 & 371 & 318 & 0.98
  & 340 & 272 & 1.03 & 69 & 296 & 323 & 300 & 439 \\
SCDM         &  6.9 & 381 & 334 & 4.4 & 349 & 290 & 375 & 325 & 0.58
  & 342 & 278 & 0.60 & 92 & 308 & 318 & 307 & 425 \\ 
$\tau$CDM    &  6.9 & 509 & 464 & 4.4 & 485 & 430 & 464 & 412 & 0.57
  & 437 & 371 & 0.58 & 70 & 392 & 399 & 398 & 535 \\
\end{tabular}
\end{center}
\end{table*}

\begin{figure*}
\begin{center}
\epsfig{file=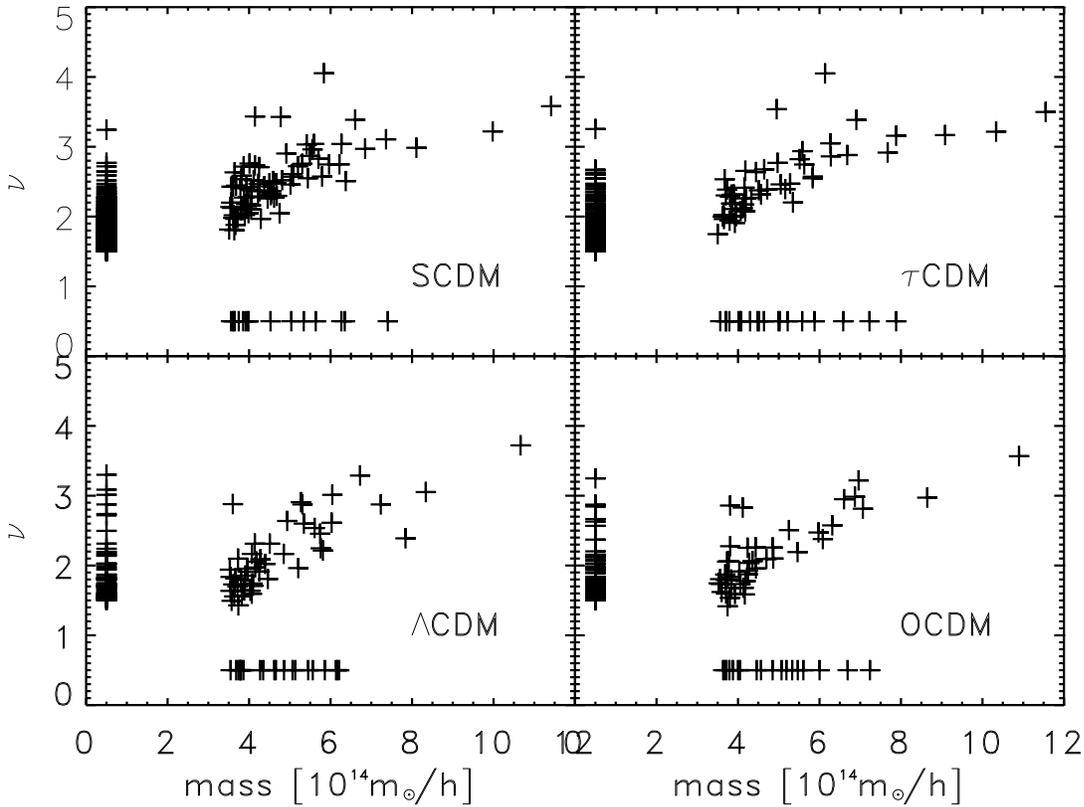,width=130mm}
\end{center}
\vspace{1.0cm}
\caption{The mass of the clusters in our simulations
against the height of the corresponding peaks in the initial
conditions, once these are smoothed with a top-hat with the 
characteristic radius listed in Table 2. All clusters with mass greater
than $3.5\times 10^{14}\,h^{-1}\,M_{\odot}$ 
and all peaks with height greater than $\nu=1.5$ are shown. There
are 351, 239, 84, and 83 unmatched peaks in the SCDM, $\tau$CDM,
$\Lambda$CDM, and OCDM model, respectively.} 
\label{peaks_mass}
\end{figure*}

\subsection{The Selection of Peaks}
\label{pselect}

We identify peaks in the initial conditions of the simulations by
binning up the initial particle distribution on a $128^3$ mesh 
using a cloud--in--cell (CIC) assignment and
then smoothing with a gaussian or a top-hat with
characteristic scale $R$ corresponding to $M_{min}=3.5\times 10^{14} 
h^{-1}\mbox{M}_{\odot}$. A peak is then taken to be any
grid point at which the smoothed density is greater than that of its
26 nearest neighbours. The dimensionless height of a peak, $\nu$, 
is defined by dividing its overdensity by the {\it rms} overdensity,
$\Delta$, which we list in Table 2. Again, within the matched set 
there is a close correspondance between the peaks found in the four 
models. In addition, the peaks found with gaussian smoothing 
correspond closely to those found with top-hat smoothing. 

Particle peculiar velocities are binned up and smoothed in an 
identical way and the peculiar velocity of a peak is taken to be the 
value at the corresponding grid point. In Table 2 we list the {\it
rms} peculiar velocity of the peaks found in each model. Again this
is scaled up to the value expected at $z=0$ according to linear
theory. It differs slightly from the value predicted by inserting 
the power spectrum of the simulation directly into equation (13) because
there are realisation to realisation fluctuations depending on the {\it 
phases} of the Fourier components. As it should, the {\it rms} 
peculiar velocity averaged over all grid points agrees very well 
with the value found by putting the simulation power spectrum into
equation (11).

%%%%%%%%%%%%%%%%%%%%%%%%%%%%%%%%%%%%%%%%%%%%%%%%%%%%%%%%%%%%%%%%%%%%%%%%%%%%%%%%

\subsection{The Selection of Clusters}
\label{selection}

We define clusters in our simulations in the same way as 
White et al. (1993). High-density regions at $z=0$  are located using a 
friends-of-friends group finder with a small linking length 
(b=0.05), and their
barycentres are considered as candidate cluster centres. Any candidate
centre for which the mass within $1.5\,h^{-1}$\,Mpc 
exceeds $M_{min}$ is identified as a candidate cluster. The final cluster
list is obtained by deleting the lower mass candidate in all
pairs separated by less than $1.5\,h^{-1}$\,Mpc. In the following we will normally 
consider only clusters more massive than $M_{min}=3.5\times 10^{14} 
h^{-1}\mbox{M}_{\odot}$. The number of clusters found in each
simulation is listed in Table 2. As already noted, the individual 
clusters in the different simulations of the matched set correspond 
closely. Despite the normalisation to cluster abundance it appears
as though the SCDM model has significantly more clusters than the
others. This is a reflection of its steeper power spectrum together
with the value of $M_{min}$ we have chosen. For $M_{min}=5.5\times
10^{14}h^{-1}\mbox{M}_{\odot}$ all the models have about 20 clusters.

We define the peculiar velocity of each cluster at $z=0$ to be the
mean peculiar velocity of all the particles within the
$1.5\,h^{-1}$\,Mpc sphere. The peculiar velocity of the cluster at
earlier times is taken to be the mean peculiar velocity of these
particles. Consistent with this, we define the position of the
cluster at each time to be the barycentre of this set of particles. At
$z=0$ this is very close to, but not identical with the cluster centre
as defined above. We give the {\it rms} values of the initial (linear) and
final ($z=0$) peculiar velocities of the clusters in each of our
models in Table 2. The initial values have been scaled up to
the linear values predicted at $z=0$. It is clear
that these substantially underestimate the actual values, a result we
discuss in more detail below. We note that the present-day
properties of clusters in these simulations are considered in
much more detail in Thomas et al.\ (1998).

%%%%%%%%%%%%%%%%%%%%%%%%%%%%%%%%%%%%%%%%%%%%%%%%%%%%%%%%%%%%%%%%%%%%%%%%%%%%%%%%

\section{Comparison of the Peak Model with Simulations}
\label{Results}

\subsection{The Cluster-Peak Connection}
\label{clusterpeak}

The extent to which dark haloes can be associated with peaks of the
smoothed initial density field is somewhat controversial. Frenk et
al. (1988) concluded that, for appropriate choices of filter scale
and peak height, the correspondance is good, whereas Katz
et al. (1993) claimed that ``there are many groups of high mass that
are not associated with any peak''. The result of correlating the
peaks in the initial conditions of our simulations with the initial
positions of our clusters is illustrated in Fig.~1. 
We consider a peak and a cluster to be associated if their separation
is less than $4\,h^{-1}$\,Mpc (comoving). We find that
the barycenters of 70\% and 80\% of the clusters with masses exceeding 
$3.5 \times 10^{14}\,h^{-1}\,M_{\odot}$ lie within of a peak with 
$\nu>1.5$ for the low and high $\Omega$ models, respectively. 

Fig.~1 shows that there is, as expected, a correlation between the
height of a peak and the mass of the corresponding cluster. In
addition, combining the peak heights with the $\Delta$ values from 
Table 2, we see that the extrapolated {\it linear} overdensities
of the peaks at redshift zero are similar
but somewhat larger than the threshold value of 1.69 used in the
standard Press-Schechter approach to analysing structure formation.

%%%%%%%%%%%%%%%%%%%%%%%%%%%%%%%%%%%%%%%%%%%%%%%%%%%%%%%%%%%%%%%%%%%%%%%%%%%%%%%%

\subsection{Linear Peculiar Velocities of Peaks and Clusters}
\label{peakclusvel}

Given the excellent correspondance between peaks of
the smoothed linear density field and the initial positions of
clusters, it is natural to compare the smoothed peculiar 
velocity at a peak with the mean initial peculiar velocity of its 
associated cluster. We show such a comparison in Fig.~2, again based
on top-hat smoothing of both position and peculiar velocity fields
using the characteristic radii listed in Table 2.
All velocities are scaled up to the expected value at $z=0$
according to linear theory. The correlation is clearly excellent in all
cases, and is similar if gaussian rather than top-hat smoothing is
used. The {\it rms} difference in peculiar velocity between a cluster and its
associated peak is 16\%, 16\%, 23\%, and 17\% of the corresponding
$\sigma_p$ value listed in Table 2 for the OCDM, $\Lambda$CDM,
SCDM and $\tau$CDM simulations respectively. The somewhat larger 
percentage for the SCDM model is probably a consequence of the greater
influence of small-scale power in this case.

\begin{figure*}
\begin{center}
\epsfig{file=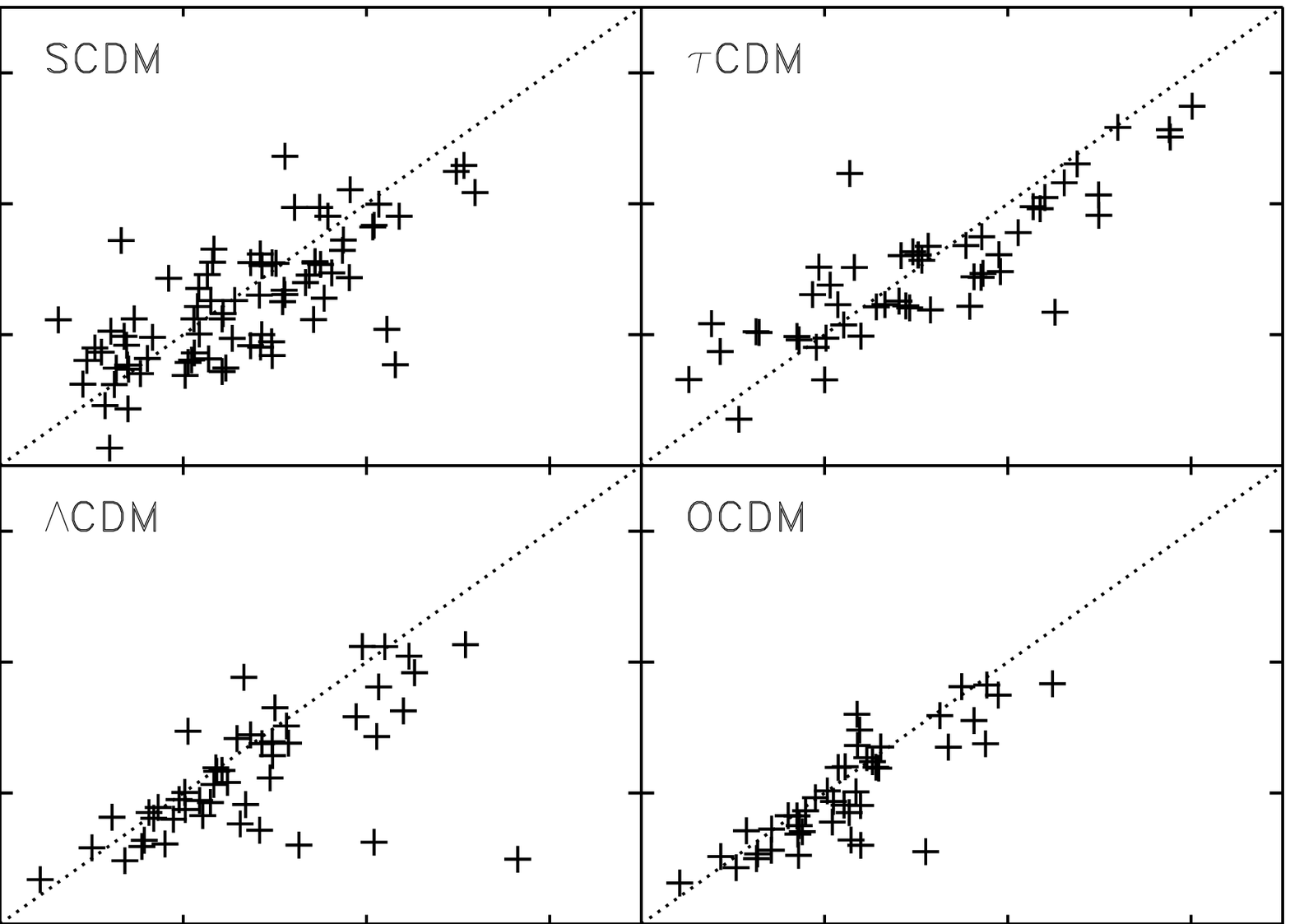,width=130mm}
\end{center}
\vspace{1.cm}
\caption{The initial peculiar velocities of clusters in each of
our four cosmogonies are compared to the linear peculiar velocities of 
their associated peaks. The linear peculiar velocity field was
smoothed with a top--hat in the same way as the density field
in order to obtain the peak peculiar velocities.}
\label{peaks_clus_vel}
\end{figure*}

%%%%%%%%%%%%%%%%%%%%%%%%%%%%%%%%%%%%%%%%%%%%%%%%%%%%%%%%%%%%%%%%%%%
%
\subsection{The Growth of Cluster Peculiar Velocities}
\label{scalecluster}

If cluster peculiar velocities grew according to linear theory
the scaled initial velocities discussed in the last section and
plotted in Fig.~2 would correspond to the actual velocities of the
clusters at $z=0$. In Fig.~3 we show scatter diagrams in which these
two velocities are plotted against each other.
It is evident that in fact the agreement is quite
poor and that there is a systematic trend for the true cluster
velocity to be larger than the extrapolated linear value. This is
reflected in the substantial difference between the {\it rms} 
values of these two quantities listed in Table 2. It is 
presumably a consequence of nonlinear gravitational forces
accelerating the clusters.

\begin{figure*}
\begin{center}
\epsfig{file=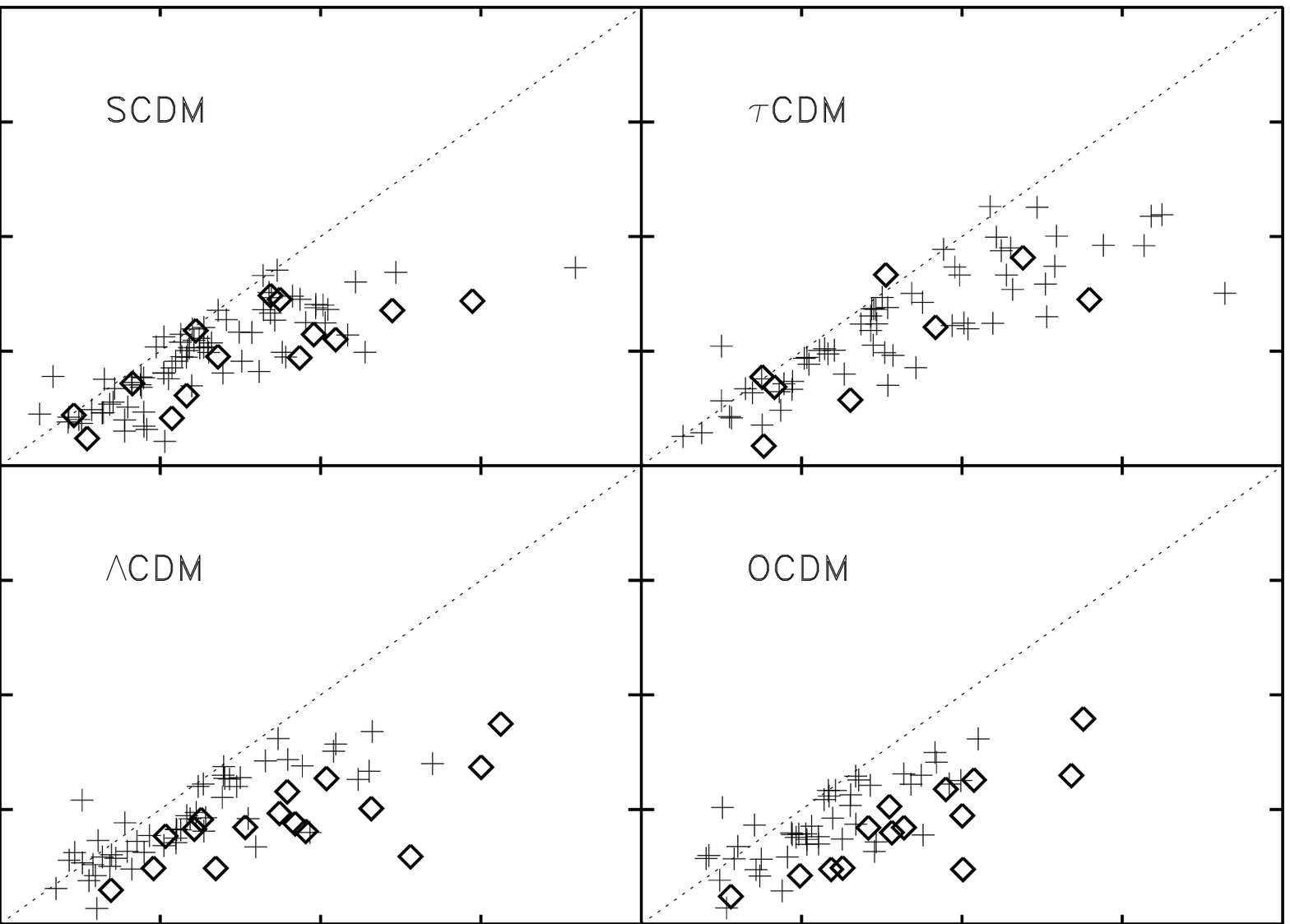,width=130mm}
\end{center}
\vspace{1.cm}
\caption{The initial peculiar velocities of clusters in each of
our four cosmogonies, scaled up to $z=0$ using linear theory, are 
compared to their actual peculiar velocities at $z=0$. Diamonds denote
clusters which have a neighbour within $10\,h^{-1}$Mpc while crosses
denote more isolated clusters.}
\label{clus_clus_vel}
\end{figure*}

Some confirmation of this is provided by Fig.~4 where we plot the
peculiar velocity in units of its initial value for five clusters 
from each of our cosmologies. At early times the
peculiar velocities all grow as expected from linear theory (indicated
in the figures by a dotted line) but at later times the behaviour is
more erratic and most clusters finish with larger velocities
than predicted.

Further evidence that late--time nonlinear effects are responsible for 
this discrepancy comes from Fig.~3. In this plot all clusters that
have a neighbour within 10\,$h^{-1}$Mpc are indicated with a diamond
while more isolated clusters are indicated by a cross. It is evident
that deviations from linear theory are substantially larger for the
``supercluster'' objects than for the rest. These objects also have 
systematically larger peculiar velocities at $z=0$. Their {\it rms}
peculiar velocity is around 20 to 30\% larger than that of the sample as
a whole.

For the $\tau$CDM model, we have run a second realization of the
power spectrum. We have extracted a cluster sample in the same fashion
as described above. The {\it rms\/} peculiar velocity of the 
clusters at $z=0$ is $\sigma_{z=0}=511$\,km/sec. The extrapolated
{\it rms} linear peculiar velocity is $\sigma_{z=0}=394$\,km/sec.
These numbers are very close to the values obtained for the first
realization. Although two simulations are not a good statistical
sample, we conclude that there is no realization dependence
of the mis--match between the extrapolated linear and the actual
peculiar velocities of galaxy clusters.

It might be thought that this anomalous acceleration of clusters at
late times was a consequence of the relatively small radius,
$1.5h^{-1}$Mpc, which we use to define our clusters. Material
could, perhaps be ejected asymmetrically from this region during the
merging events by which clusters form. We have searched for such
effects by redefining clusters to be all the material contained within
a radius of 3 or $5\,h^{-1}\,$Mpc and then repeating the analysis for the
same set of objects as before. In most cases this turned out to make 
very little difference to either the initial or the final velocities
measured, and it did nothing to reduce the
discrepancy between them. The relevant nonlinear effects are
acting on significantly larger scales. We repeated this procedure
going as far out as $25\,h^{-1}\,$Mpc from the cluster center. At a 
radius of $10\,h^{-1}\,$Mpc, the difference between the {\it rms\/} 
peculiar velocity and the extrapolated {\it rms} linear peculiar 
velocity is only 10\%. By a radius of $20\,h^{-1}\,$Mpc, the numbers 
have finally converged.
\begin{figure*}
\begin{center}
\epsfig{file=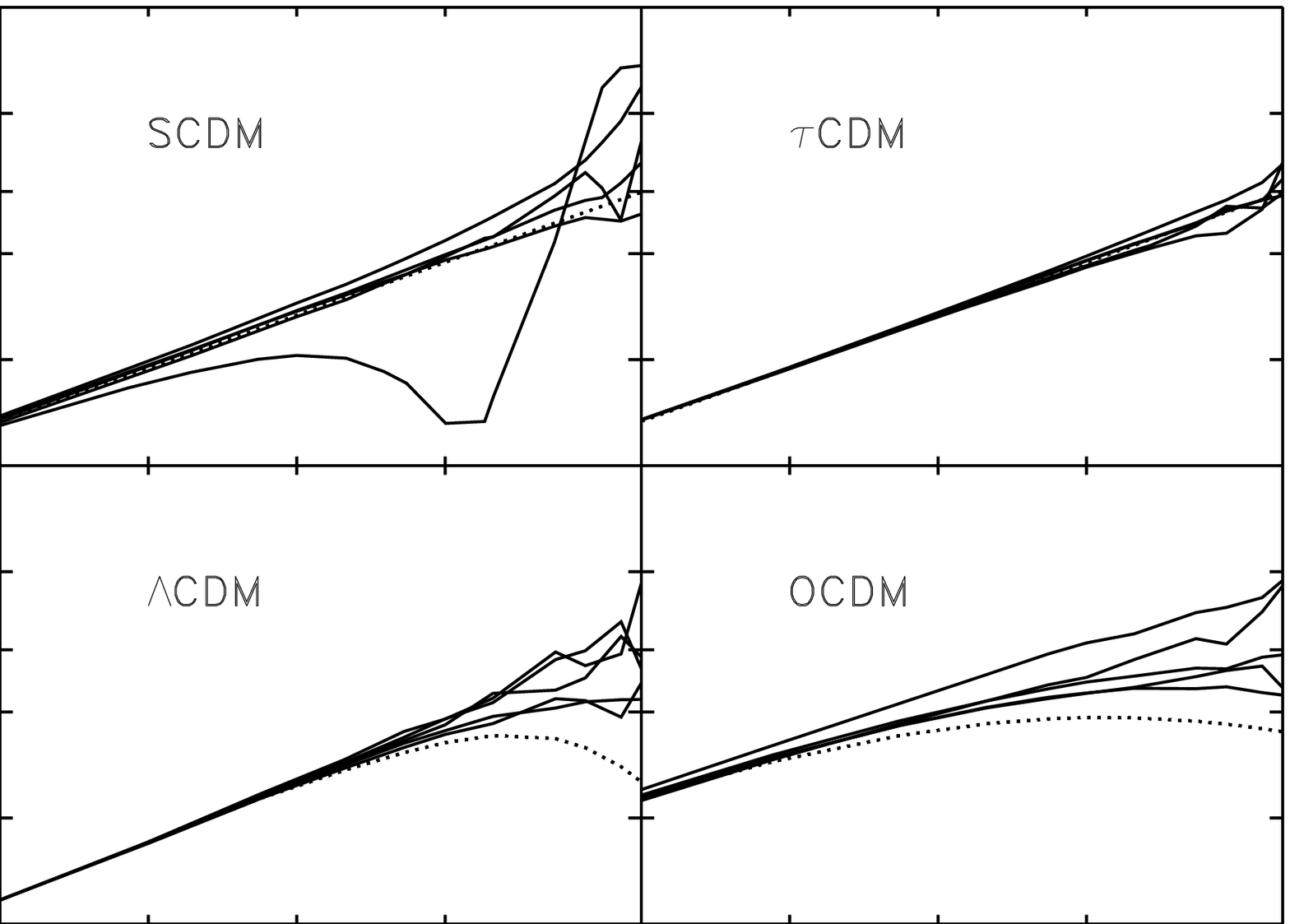,width=130mm}
\end{center}
\vspace{1.cm}
\caption{The evolution with expansion factor $a$ of the ratio 
$|\vec{v}(a)|/|\vec{v}_0|$ for five clusters from each of our
four cosmogonies (solid lines) is compared with the evolution
predicted by linear theory (dotted line). In some of the cases,
merging leads to abrupt changes in this ratio -- the most
impressive case can be seen for one of the SCDM clusters.}
\label{pv_all}
\end{figure*}

The discrepancy
between the {\it rms\/} peculiar velocity of clusters  
and their extrapolated {\it rms} linear peculiar velocity is
independent of any smooting of the density field. 
With our choice of smoothing filter, the linear peculiar velocities
of our clusters match those of their associated peaks as
well as the {\em rms\/} value predicted by linear theory 
when the simulated
realization of the power spectrum and the proper expression
for the peculiar velocities (eq.\ 13) is used. Previous
work (e.g.\ Borgani et al.\ 1997) has tried to match N--body 
data with linear theory by tuning the filter scale.
Our results undermine the physical basis for such 
procedure.

%%%%%%%%%%%%%%%%%%%%%%%%%%%%%%%%%%%%%%%%%%%%%%%%%%%%%%%%%%%%%%

\section{Conclusions}
\label{summ}

We have investigated the peculiar velocities predicted for 
galaxy clusters by theories in the Cold Dark Matter
family. A widely used hypothesis identifies rich clusters with high
peaks of a smoothed version of the linear density fluctuation
field. Their peculiar velocities are then obtained by extrapolating
the similarly smoothed linear peculiar velocities at the
positions of these peaks. We have tested this 
using a set of four large high--resolution N--body 
simulations. We identify galaxy clusters at $z=0$
and then trace the particles they consist of back to earlier
times. In the initial density field,
the barycenters of 70\% and 80\% of the clusters with masses exceeding 
$3.5 \times 10^{14}\,h^{-1}\,M_{\odot}$ lie within 
$4\,h^{-1}$\,Mpc (comoving) of a peak with 
$\nu>1.5$ for the low and high $\Omega$ models,
respectively. Furthermore, the mean linear peculiar velocity of the material
which forms a cluster at $z=0$ agrees well with the value at
that peak.

However, the late--time growth of peculiar velocities is
systematically underestimated by linear theory. At the time
clusters are identified, i.e.\ at $z=0$, we find that the {\it rms}
peculiar velocity is about 40\% larger than predicted. Nonlinear
effects are particularly important in superclusters; the
{\it rms} values for clusters which are members of superclusters
are about 20\% to 30\% larger than those for isolated clusters.

%%%%%%%%%%%%%%%%%%%%%%%%%%%%%%%%%%%%%%%%%%%%%%%%%%%%%%%%%%%%%%

\section*{Acknowledgements}

The simulations were carried out on the Cray T3D's and T3E's at the
Computer Center of the Max--Planck--Gesellschaft in Garching
and at the Edinburgh Parallel Computing Centre. Postprocessing was done on the
IBM SP2 at the Computer Center of the Max--Planck--Gesellschaft 
in Garching. We thank George Efstathiou and John Peacock for valuable
comments. JMC would like to thank Matthias Bartelmann, Antonaldo
Diaferio, Adi Nusser, Ravi Sheth, and Mirt Gramann
for numerous helpful and interesting discussions, 
and Volker Springel for providing his smoothing code.

This work was supported in part by the EU's TMR Network for 
Galaxy Formation. CSF acknowledges a PPARC Senior Research
Fellowship.


\begin{thebibliography}{}
\bibitem[\protect\citename{Bahcall }{1996}]{Bahcall96}
Bahcall N.A., Oh S.P., ApJ, 462L, 49B (1996)
\bibitem[\protect\citename{Bardeen {\em et al.\/} }{1986}]{Bardeen86}
Bardeen J.M., Bond J.R., Kaiser N., Szalay A.S., ApJ, 304, 15 (1986) (BBKS)
\bibitem[\protect\citename{Bharadwaj \& Sethi }{1998}]{Bharadwaj98} 
Bharadwaj S., Sethi S.K., ApJS, 114, 37B (1998)
\bibitem[\protect\citename{Bond \& Efstathiou }{1984}]{Bond84}
Bond, J.R., Efstathiou, G., ApJ, 285, L45 (1984)
\bibitem[\protect\citename{Borgani et al. }{1997}]{Borgani97}
Borgani S., Da Costa L.N., Freudling W., Giovanelli R., Haynes M.P.,
Salzer J., Wegner G., ApJ, 482, L121 (1997)
\bibitem[\protect\citename{Bond }{1992}]{Carroll92}
Carroll S.M., Press W.H., Turner E.L., ARAA, 30, 499 (1992)
\bibitem[\protect\citename{Couchman }{1995}]{Couchman95}
Couchman H.M.P., Thomas P.A., Pearce F.R., ApJ, 452, 797 (1995)
\bibitem[\protect\citename{Croft \& Efstathiou }{1994}]{Croft94}
Croft R.A.C., Efstathiou G., MNRAS, 268, L23 (1994)
\bibitem[\protect\citename{Efstathiou {\em et al.\/} }{1985}]{Efstathiou85}
Efstathiou G., Davis M., Frenk C.S., White S.D.M., ApJS, 57, 241 (1985)
\bibitem[\protect\citename{Eisenstein  }{1997}]{Eisenstein97}
Eisenstein D.J., astro-ph/9709054
\bibitem[\protect\citename{Eke {\em et al.\/} }{1996}]{Eke96}
Eke V.R., Cole S., Frenk C.S., MNRAS, 282, 263 (1996)
\bibitem[\protect\citename{Frenk {\em et al.\/} }{1988}]{Frenk88}
Frenk C.S., White S.D.M., Davis M., Efstathiou G., ApJ, 327, 507 (1988)
\bibitem[\protect\citename{Heath }{1977}]{Heath77}
Heath D.J., MNRAS, 176, 1 (1977)
\bibitem[\protect\citename{Jenkins et al. }{1996}]{Jenkins96}
Jenkins A., Frenk C.S., Pearce F.R., Thomas P.A., 
Hutchings, Colberg J.M., White S.D.M., Couchman H.M.P.,
Peacock J.A., Efstathiou G.P., Nelson A.H., ''The VIRGO
Consortium: Simulations of Dark Matter and Galaxy Clustering'',
in Dark and Visible Matter in Galaxies and Cosmological
Implications, ed.s Persic M., Salucci P., ASP Conference
Series (1996)
\bibitem[\protect\citename{Jenkins et al. }{1998}]{Jenkins98}
Jenkins A., Frenk C.S., Pearce F.R., Thomas P.A., 
Colberg J.M., White S.D.M., Couchman H.M.P.,
Peacock J.A., Efstathiou G.P., Nelson A.H. (The Virgo Consortium),
ApJ, 499, in press (1998)
\bibitem[\protect\citename{Katz {\em et al.\/} }{1993}]{Katz93}
Katz N., Quinn T., Gelb J.M., MNRAS, 265, 689 (1993) (KQG)
\bibitem[\protect\citename{Lahav }{1991}]{Lahav91}
Lahav O., Lilje P.B., Primack J.R., Rees M.J., MNRAS, 251, 128 (1991)
\bibitem[\protect\citename{MacFarland }{1997}]{MacFarland97}
MacFarland T., Pichlmaier J., Pearce F.R., Couchman H.M.P.,
A Parallel P3M Code for Very Large
Scale Cosmological Simulations, presented at
the European Cray-SGI MPP Workshop, Paris 1997,
\protect{http://www.rzg.mpg.de/$\sim$tmf/package/paper.html}
(only Internet version available)
\bibitem[\protect\citename{Monaco }{1998}]{Monaco98}
Monaco P., Fund.\ Cosm.\ Phys., in press (1998)
\bibitem[\protect\citename{Pearce et al. }{1995}]{Pearce95}
Pearce F.R., Couchman H.M.P., Jenkins A.R., Thomas P.A., 
'Hydra -- Resolving a parallel nightmare', in 
Dynamic Load Balancing on MPP systems
\bibitem[\protect\citename{Pearce }{1997}]{Pearce97}
Pearce, F.R., Couchman, H.M.P., ''Hydra: A parallel adaptive
grid code'', submitted to New Astronomy (1997)
\bibitem[\protect\citename{Peebles }{1993}]{Peebles93}
Peebles, P.J.E., Principles of Physical Cosmology,
Princeton University Press 1993
\bibitem[\protect\citename{Strauss }{1995}]{Strauss95}
Strauss, M.A., Willick, J.A., Phys.\ Rep., 261, 271 (1995)
\bibitem[\protect\citename{Suhiyama }{1995}]{Sugiyama95}
Sugiyama N., ApJS, 100, 281 (1995)
\bibitem[\protect\citename{Thomas }{1997}]{Thomas97}
Thomas P.A., Colberg J.M., Couchman H.M.P., Efstathiou G., 
Frenk C.S., Jenkins A.R., Nelson A.H., Hutchings R.M., 
Peacock J.A., Pearce F.R., White S.D.M. (The Virgo Consortium)
accepted by MNRAS, astro-ph/9707018
\bibitem[\protect\citename{White et al. }{1995}]{White95}
White M., Gelmini G., Silk J., Phys. Rev. D, 51, 2669 (1995)
\bibitem[\protect\citename{White {\em et al.\/} }{1993}]{White93}
White S.D.M., Efstathiou G., Frenk C.S., MNRAS, 262, 1023 (1993)
\end{thebibliography}
\end{document}